\documentclass[aps,prb,twocolumn, 10pt, superscriptaddress]{revtex4-2}
\usepackage{graphicx}
\usepackage{amsmath}
\usepackage{amssymb}
\usepackage{mathtools}
\usepackage{xcolor}
\usepackage[colorlinks=true, citecolor=blue, urlcolor=blue, linkcolor=blue,bookmarks=false,hypertexnames=true]{hyperref} 
\usepackage[T1]{fontenc} 
\usepackage{braket}

\begin{document}
\graphicspath{}

\title{Probing the zero energy shell wave functions of triangular graphene quantum dots with broken sublattice symmetry using a localized impurity}

\author{Alina Wania Rodrigues} \thanks{awaniaro@uottawa.ca}
\affiliation{Department of Physics, University of Ottawa, Ottawa, Ontario, K1N 6N5, Canada}

\author{Daniel Miravet} \thanks{dmiravet@uottawa.ca}
\affiliation{Department of Physics, University of Ottawa, Ottawa, Ontario, K1N 6N5, Canada}

\author{James Lawrence}
\affiliation{Department of Chemistry, National University of Singapore, 117543 Singapore}

\author{Jiong Lu}
\affiliation{Institute for Functional Intelligent Materials, National University of Singapore, 117544 Singapore}
\affiliation{Department of Chemistry, National University of Singapore, 117543 Singapore}

\author{Pawe\l\ Hawrylak}
\affiliation{Department of Physics, University of Ottawa, Ottawa, Ontario, K1N 6N5, Canada}

\date{\today}

\begin{abstract}

We present here a method of probing the wave functions  of a degenerate shell in a triangular graphene quantum dot, triangulene, using a localized substitutional impurity. We demonstrate its applicability to the example of aza-triangulenes. Using the analytical solution for degenerate states of an all-carbon triangulene as a basis for a triangulene containing a nitrogen impurity, we predict the structure of the zero energy shell in the presence of this impurity. We show that the impurity allows probing of the wave functions of a degenerate shell on a carbon site where it is located. We confirm our predictions by a comparison with the tight-binding and \textit{ab-initio} calculation as well as with experiment.

\end{abstract}

\maketitle

\section{Introduction}
Synthetic correlated electron systems offer the possibility to design topological quantum matter with properties vastly different from its constituents, with superconductivity, ferromagnetism and incompressible liquids of the fractional quantum Hall effect as good examples. Electronic correlations and entanglement associated with it originate from electron-electron (e-e) interactions in a flat band, i.e. degenerate electronic shells. A flat band is realized in the lowest degenerate Landau level of a two dimensional electron gas \cite{Stormer_Gossard_1999, Halperin_Read_1993, Du_West_1993, Willett_Pfeiffer_1993, de-Picciotto_Mahalu_1997, Saminadayar_Etienne_1997, Byszewski_Hawrylak_2006}, in  twisted  graphene layers \cite{Bistritzer_MacDonald_2011, Cao_Jarillo-Herrero_2018a, Cao_Jarillo-Herrero_2018b} or by on surface synthesis of triangulenes \cite{Mishra_Fasel_2020, Pavlicek_Gross_2017, Guclu_Hawrylak_2009, Su_Lu_2019, Su_Lu_2020, Su_Lu_2021, Lawrence_Lu_2023, Valenta_Juríček_2022}, triangular graphene quantum dots with broken sublattice symmetry.

Triangulenes are graphene nanostructures fabricated with atomic precision through on-surface synthesis, allowing for the design of the degenerate electronic shell at the Dirac point \cite{Mishra_Fasel_2020, Pavlicek_Gross_2017, Guclu_Hawrylak_2009, GQD_book_2014, Mishra_Ruffieux_2019, Mishra_Ruffieux_2021, Su_Lu_2019, Su_Lu_2020, Su_Lu_2021, Turco_Ruffieux_2023, Lawrence_Lu_2023, Valenta_Juríček_2022}. They consist of carbon atoms arranged on a hexagonal lattice, shaped as an equilateral triangle, and terminated with a zig-zag edge. They have been investigated in  recent years due to the development of the on-surface synthesis of triangulenes of different sizes and in different configurations. Such experiments were previously inaccessible due to the highly reactive nature of these structures.

The presence of the zig-zag edge leads to triangulene exhibiting a shell of degenerate states around the Fermi energy, with the number growing linearly with the system size. These properties have first been demonstrated on the single particle level through analytical and tight-binding calculations \cite{Fernandez-Rossier_Palacios_PRL_2007, Potasz_Hawrylak_PRB_2010}, but have since been confirmed by mean field approaches, such as \textit{ab-initio} and Hartree Fock calculations, as well as by exact diagonalization results \cite{Guclu_Hawrylak_PRL_2009, Guclu_Hawrylak_PRB_2010, Potasz_Hawrylak_2012, Voznyy_Hawrylak_PRB_2011}. 

Magnetic properties of triangulenes can be inferred  using  Lieb's theorem for the bi-partite Hubbard model that predicts the total spin of the ground state corresponding to the difference between the number of A and B atoms on the bipartite graphene lattice \cite{Lieb_PRL_1989}.

In this work, we focus on a special class of triangulenes, namely aza-triangulenes, in which one of the carbon atoms has been replaced by nitrogen \cite{Wang_Pascual_Oteyza_2022, Wei_Wu_2022, Henriques_Fernández-Rossier_2023}. For example, aza-[5]triangulenes, with 5 benzene rings in a given edge, consisting of 46 atoms, have been recently realized experimentally \cite{Lawrence_Lu_2023, Vilas-Varela_Pascual_2023}. 
Replacing a carbon atom with a nitrogen atom introduces an extra electron and proton. This substitution removes an unpaired electron from one sublattice, thereby altering the total spin of the system.
In this work, we will focus on the case where nitrogen replaces a carbon atom in the majority sublattice of the triangular-shaped graphene quantum dots with zigzag edges.

We discuss here an exact analytical solution to the zero energy shell and show how the nitrogen impurity allows us to probe its wavefunctions. The theoretical predictions are compared with \textit{ab-initio} calculations and with scanning tunneling microscopy (STM) experiments.

The paper is organized as follows. We start with the introduction, followed by the rederivation of the states of the zero energy shell of a triangulene. Next, we derive the zero-energy shell of aza-triangulene and show how to extract the zero energy wavefunctions. Finally, we compare our analytical predictions with tight-binding, \textit{ab-initio} and experimental results. 

\section{The electronic states of the zero energy shell}
\label{s2}
\begin{figure}
    \centering
    \includegraphics[scale=0.35]{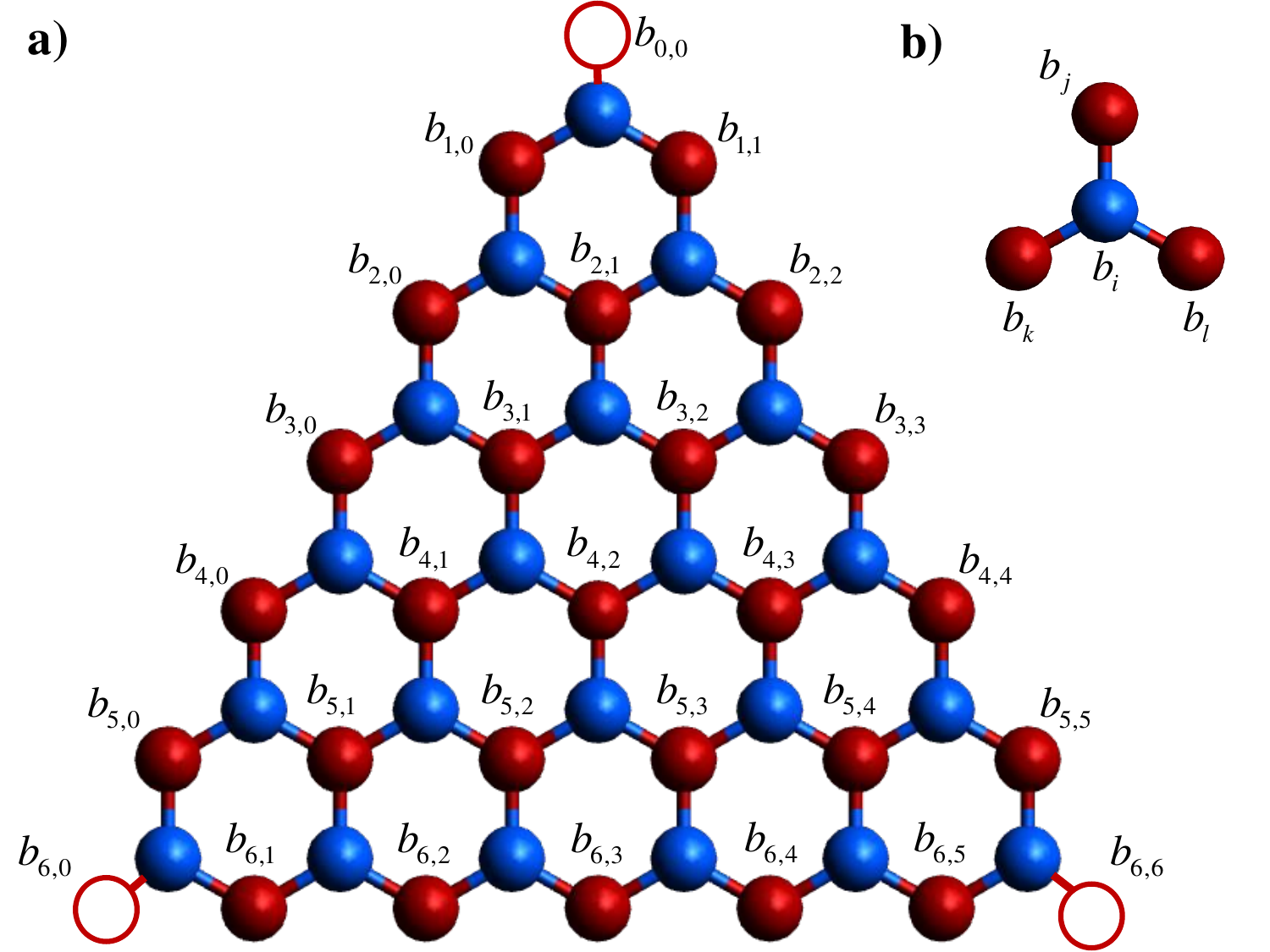}
    \caption{(a) Geometry of [5]-triangulene structure. A-atoms are plotted as red dots, B-atoms as blue dots. Empty circles in the corners of the triangle are fictitious atoms added to ensure proper boundary conditions. All A-atoms are labeled with $b_{n,m}$ coefficients. (b) B-atom with its three nearest neighbors. All atoms are labeled with their $b_i$ coefficients.}
    \label{fig1}
\end{figure}

Following Ref. \cite{Potasz_Hawrylak_PRB_2010} we derive here analytical expressions for wavefunctions of the degenerate zero energy shell of triangulene.  We consider a triangulene of an arbitrary size, consisting of two inequivalent sublattices - A and B, depicted by red and blue dots respectively in Fig. \ref{fig1}(a) for a case of [5]triangulene. 
We model this structure using the single $p_z$, nearest-neighbour tight-binding model. The Hamiltonian has the form:
\begin{equation}
\label{Eq1}
    H=t\sum_{\langle i,j \rangle} c_i^\dagger c_j,
\end{equation}
where $c_i^\dagger$ ($c_i$) are the creation (annihilation) operators of an electron on the $\rm{i^{th}}$ site, $\langle i,j \rangle$ indicate summation over the nearest neighbors, and $t$ is the hopping integral. The zero-energy shell consists of states with zero energy i.e. all states which satisfy the singular eigenvalue problem:
\begin{equation}
\label{eq2}
    H\Psi = 0.
\end{equation}
Here $\Psi$ is the wave function of the whole system and can be written as a sum of contributions coming from both sublattices: $\Psi=\Psi^A + \Psi^B$, where:
\begin{equation}
    \Psi_{\rm{A(B)}}^\alpha = \sum b_i^\alpha \phi_i^{\rm{A(B)}}.
\label{wf}
\end{equation} 
Here $\phi_i^{\rm{A(B)}}$ refer to $p_z$ orbitals localized on the $\rm{i^{th}}$ A (B) atom and coefficients $b_i^\alpha$, to be determined, specify the eigenstate $\alpha$. Let's consider an $\rm{i^{th}}$ B-atom and its three nearest neighbours (Fig. \ref{fig1}(b)). Using Eq. \ref{eq2} and projecting on $\phi_i^B$, we obtain:
\begin{equation}
\begin{split}
    b_i\bra{\phi_i^B}H\ket{\phi_i^B} + b_j\bra{\phi_i^B}H\ket{\phi_j^A} +
    b_k\bra{\phi_i^B}H\ket{\phi_k^A} + \\
    b_l\bra{\phi_i^B}H\ket{\phi_l^A} = 0.
\end{split}
\label{eq3}
\end{equation}
Using the expressions $\bra{\phi_i^B}H\ket{\phi_j^A} = \bra{\phi_i^B}H\ket{\phi_k^A} = \bra{\phi_i^B}H\ket{\phi_l^A} =t $ and $\bra{\phi_i^B}H\ket{\phi_i^B} = 0$, we can write Eq. \ref{eq3} as condition:
\begin{equation}
\label{sum_ober_bi}
    b_j + b_k + b_l = 0.
\end{equation} 
This means that in the nearest neighbor approximation, the sum of coefficients of the zero-energy states around each site must vanish. We will now focus our analysis on the majority sublattice - A, depicted by red dots in Fig \ref{fig1}.

Each atom is labeled by two integer numbers $n$ and $m$ which correspond to the row and column index, respectively, following the convention of Pascal's triangle. We ensure proper boundary conditions by adding auxiliary atoms in the corners of 
our triangle, depicted by empty red circles in Fig \ref{fig1}(a). We now can use Eq. \ref{sum_ober_bi} to express all coefficients $b_{n,m}$ as linear combinations of the coefficients on the left edge, i.e. $b_{n,0}$. Starting from the top of our triangle, we can obtain $b_{0,1}=-(b_{0,0}+b_{1,1})$ and $b_{2,1}=-(b_{1,0}+b_{2,0})$. Using these expressions, and Eq. \ref{sum_ober_bi} again, we can apply the procedure to the coefficients in the second row. Repeating these steps for each row, we obtain all the needed coefficients. We note that this procedure can be carried out for all triangles, irrespective of their size. A general expression for any $b_{n,m}$ coefficient can be written as:
\begin{equation}
    b_{n,m}=-(b_{n,m-1}+b_{n-1,m-1}).
    \label{b_nm}
\end{equation}
This recurrence relation connects the coefficient of each atom to the two neighboring atoms on its left.  

\begin{figure*}
    \centering
    \includegraphics[scale=0.5]{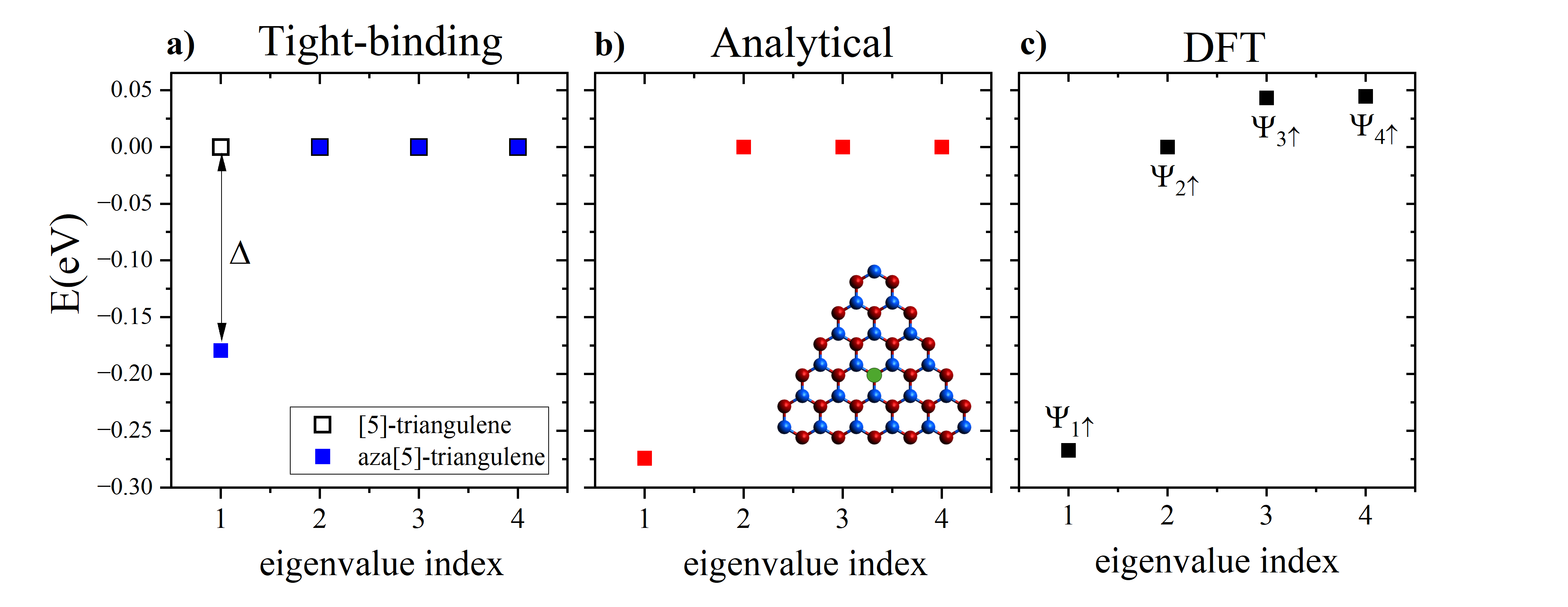}
    \caption{Degenerate shell of [5]triangulene with and without the nitrogen impurity. (a) Tight-binding calculation - black empty squares refer to [5]triangulene, while the blue squares correspond to aza-[5]triangulene. (b) Analytical result for aza-[5]triangulene, inset shows the location of the nitrogen defect in the triangulene molecule - note the green dot in the center of the triangle. For both the tight-binding and analytical calculation the on-site energy of nitrogen $\delta$ was set to $5t$. (c) DFT calculation for aza-[5]triangulene. In all three cases of aza-[5]triangulene, the impurity is localized as shown on the inset of (b).}
    \label{fig2}
\end{figure*}

We will now focus on an arbitrary atom labeled by $(k,0)$, lying on the left edge, and another atom labeled by $(n,m)$. Applying Eq. \ref{b_nm} to all coefficients on the left of atom $b_{n,m}$ leading up to the $b_{k,0}$ atom, one can notice that the sign of the calculated coefficients alternates with each progression. Specifically, when we reach atom $b_{k,0}$ on the left edge, the sign in front of the final expression becomes $(-1)^m$, where $m$ represents the number of steps needed to arrive to the edge. Additionally, the number of paths connecting these two atoms corresponds to the binomial coefficient $\begin{pmatrix} m \\ n-k \\ \end{pmatrix}$. These two observations, allow us to formulate the following expression for the coefficient $b_{n,m}$:

\begin{equation}
    b_{n,m}=(-1)^m\sum_{k \ge n-m}^n \begin{pmatrix} m \\ n-k \\ \end{pmatrix}
    b_{k,0}.
\label{coeff_def}
\end{equation}
We can now use our boundary conditions, i.e., impose vanishing of all the corner coefficients:
$b_{0,0}=b_{N+1,0}=b_{N+1,N+1}=0$. This reduces the number of independent coefficients to $N-1$, where $N$ is the number of A-atoms on a given edge.

A similar analysis can be conducted for the B-sublattice. Again, we have to ensure proper boundary conditions, so additional atoms are added in such a way that all A-atoms have three nearest neighbors. We can now repeat the steps described above, using Eq. \ref{sum_ober_bi} to reduce the number of independent coefficients. However, in this case, after imposing the vanishing of coefficients on the auxiliary atoms, we end up with only one independent coefficient, which results in a trivial solution. This leads to the conclusion that the zero energy states can only consist of coefficients from the majority sublattice, which builds up the zig-zag edge. 
Now, using Eq. \ref{wf} and Eq. \ref{coeff_def}, a general form of the states for the zero-energy shell can be written as:
\begin{equation}
\label{wf_gen}
    \Psi = \sum_{n=0}^{N+1}\sum_{m=0}^{N}\left [ (-1)^m\sum_{k \ge n-m}^n \begin{pmatrix} m \\ n-k \\ \end{pmatrix}
    b_{k,0} \right] \phi_{n,m}^A
\end{equation}
where $\phi_{n,m}^A$ is the $p_z$ orbital on the A-type site $(n,m)$. 
In this expression only the $N-1$ coefficients corresponding to atoms from the left edge are independent. Thus, we can construct $N-1$ linearly independent eigenvectors which span the subspace with zero-energy states. Our $N-1$ linearly independent eigenvectors, are in general non-orthogonal. We therefore use the Gram-Schmidt algorithm to orthogonalize them. This way an orthonormal basis can be constructed, in which the zero-energy states can be written. We note that all the above calculations have been carried out for an all-carbon triangulene. This analysis will now be extended to aza-triangulenes.

To incorporate the effect of a nitrogen impurity into our tight-binding model, we have altered the on-site energy corresponding to the nitrogen site. The analytical calculation is conducted as follows.
A new Hamiltonian, in which the nitrogen impurity is treated as a small perturbation can be constructed:
\begin{equation}
    H^{\textrm{aza}} = H + V_\textrm{imp}.
\end{equation}
Here $H$ is the original, all-carbon Hamiltonian, and $V_\textrm{imp} = \delta c^\dagger_\textrm{imp}c_\textrm{imp}$. $\delta$ is a parameter which corresponds to the on-site energy of nitrogen. The analytical solution for the degenerate shell of $H$ has already been obtained:
\begin{equation}
    H\ket{A} = \varepsilon_0\ket{A},
\end{equation}
where $\ket{A}$ are the analytical vectors obtained above, and $\varepsilon_0$ is the energy of the degenerate shell. We will now use the set $\{\ket{A}\}$ as a basis to express the states of the degenerate shell of aza-triangulene $H^{\textrm{aza},0}$:
\begin{equation}
    H^{\rm{aza},0}_{\alpha\beta} = \bra{A_\alpha}H^{\rm{aza}}\ket{A_\beta} = \bra{A_\alpha}H\ket{A_\beta} + \bra{A_\alpha}V_{\rm{imp}}\ket{A_\beta}.
\end{equation}
Since the energy of the zero-energy states is zero, the first term disappears, and we are left with:
\begin{equation}
    H^{\rm{aza},0}_{\alpha\beta} = \bra{A_\alpha}V_{\rm{imp}}\ket{A_\beta} = \delta b^\alpha_{\rm{imp}}b^\beta_{\rm{imp}}.
\end{equation}
The resulting Hamiltonian is in general a dense matrix of a dimension $N-1$, which corresponds to the number of the zero-energy states of the original system. To obtain the new energy values for an aza-triangulene we would have to diagonalize an arbitrarily large matrix. Since the original states were degenerate, any combination of them is also an eigenstate of the system. As such, we can use the first step of the Gauss elimination algorithm to rotate our analytical vectors $\ket{A}$ to a basis, where only one $b_{\rm{imp}}$ remains non-zero. This allows us to write $H^{\rm{aza},0}$ in a diagonal form straight away, with only one non-zero eigenvalue. As a result, only one state from the zero-energy shell is shifted by $\delta |b_{\rm{imp}}'|^2$, where $b_{\rm{imp}}'$ refers to the coefficient in the rotated basis.

\section{Degenerate shell of aza-[5]triangulene}

\label{s3}

\begin{figure}
    \centering
    \includegraphics[scale=0.6]{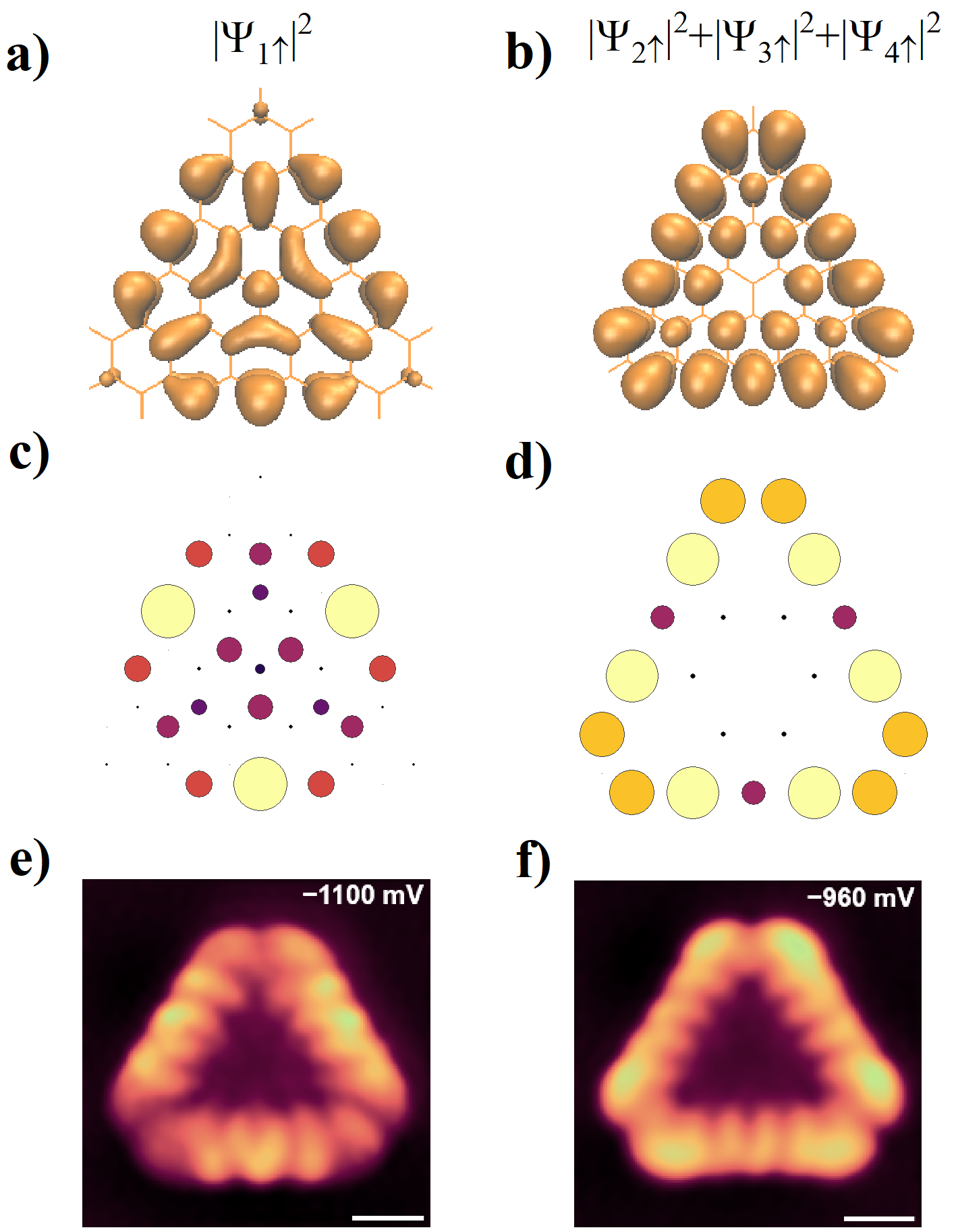}
    \caption{Degenerate shell wave function densities. The first column corresponds to the wave function density of the $\Psi_{1\uparrow}$ state marked in Fig \ref{fig2} (c), while the second column corresponds to the sum of the $\Psi_{2\uparrow}, \Psi_{3\uparrow}, \Psi_{4\uparrow}$ states, shown in Fig \ref{fig2} (c). (a) and (b) show the results of DFT calculation, and (c) and (d) show the results of analytical calculation. (e) and (f) present experimental constant current dI/dV images of aza-[5]triangulene on Au(111) \cite{Lawrence_Lu_2023}. Imaging parameters: T = 4.3 K. $I_T = 2$ nA, lock-in bias voltage oscillation amplitude = 30 mV. Scale bars = 0.5 nm. Images have been FFT-filtered to remove higher frequency electronic noise.}
    \label{fig3}
\end{figure}
\begin{figure*}
    \centering
    \includegraphics[scale=0.38]{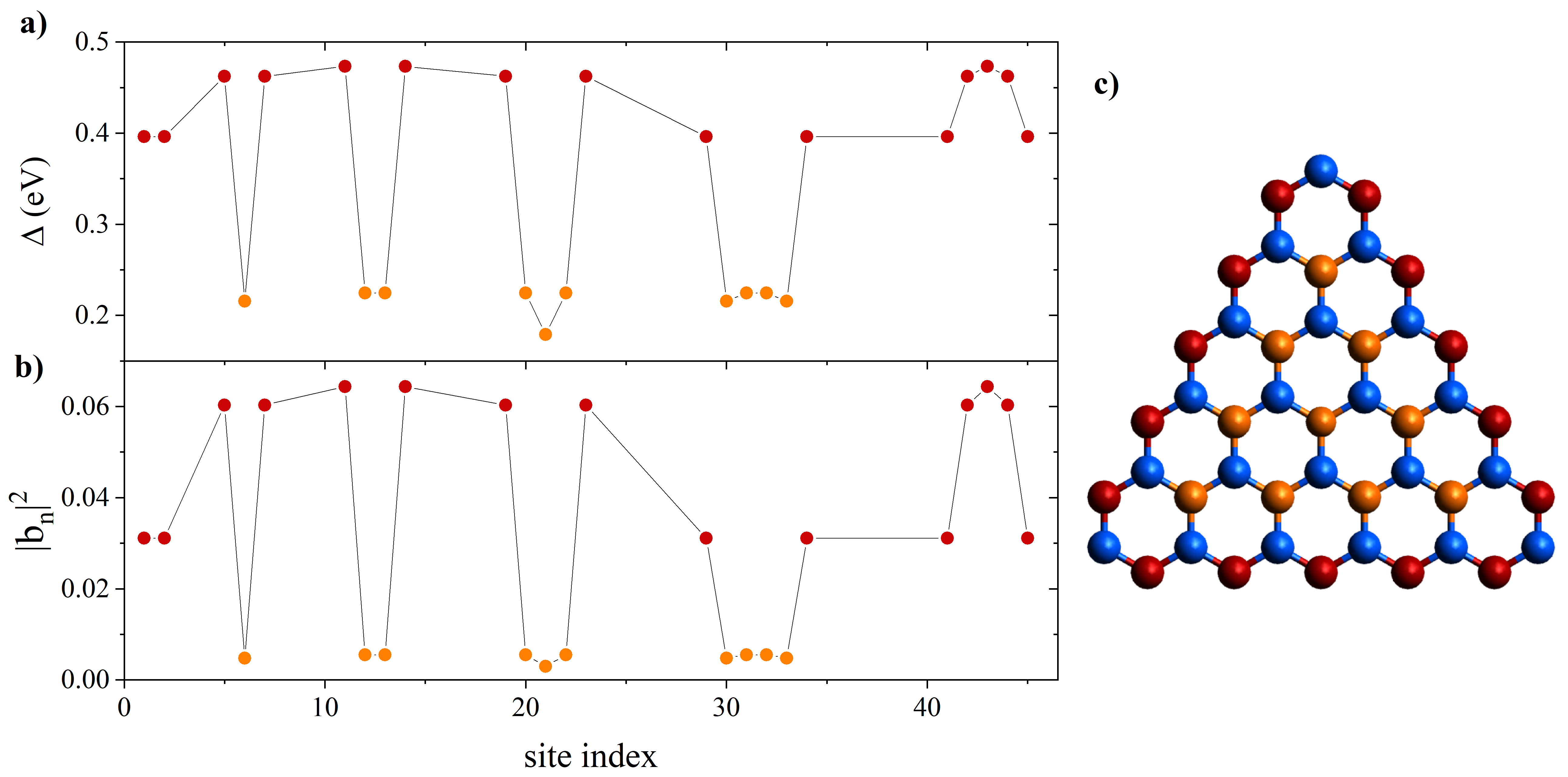}
    \caption{(a) Magnitude of the split of the degenerate shell for aza-[5]triangulene with impurity localized on different sites. Red data points correspond to the position of the impurity on red atoms shown in (c), while orange data points refer to the orange atoms in (c). (b) Magnitude of the wavefunction coefficient corresponding to the site with impurity. Red and orange data points are assigned analogically as (a). (c) Geometrical structure of aza[5]-triangulene. Edge A-atoms are marked in red, bulk A-atoms are marked in orange, while all B-atoms are marked in blue. The site index 0-45 has been allocated to each atom row by row from left to right e.g. the top blue atom has index zero, while in the second row the left red atom has index 1, and the right red one has index 2.}
    \label{fig4}
\end{figure*}

We can now apply the above procedure to a specific case of aza-[5]triangulene with a nitrogen impurity localized at the center of the system (see inset of Fig. \ref{fig2}(b)). We start by performing the tight-binding calculations. The resulting energy spectrum in the vicinity of the Fermi level has been shown in Fig \ref{fig2}(a) (blue squares) and has been compared to the degenerate shell of an all-carbon [5]triangulene (black, empty squares). One can see that the degenerate shell which previously consisted of 4 zero-energy states is preserved, with only one state having now a lower energy. The other three states remain degenerate and at zero energy. The magnitude of this split, denoted as $\Delta$, is indicated by the black arrow in Fig. \ref{fig2}(a). We compare this calculation with the analytical solution obtained through the procedure described above and presented in Fig \ref{fig2}(b). Both of these results are in a qualitative agreement. 
Discrepancies in the magnitude of the split $\Delta$ between the tight-binding and analytical solutions can be attributed to the size of the basis used. In the analytical calculation, we only considered 4 eigenvectors to span our basis. However, increasing their number narrows the difference between the analytical and tight-binding solutions. Both results fully converge when all 46 eigenvectors are used to construct the basis.
Additionally, we validated these findings through the \textit{ab-initio} DFT calculations. The resulting energy spectrum of Kohn-Sham orbitals is depicted in Fig. \ref{fig2}(c). DFT calculations were conducted within the local density approximation using the Octopus software \cite{octopus}.

Having analyzed the energy spectrum, we now move on to examine the wave function probability densities of the zero-energy shell. In this step we have compared the analytical, DFT and experimental results, all of which are presented in Fig \ref{fig3}. The experiment considered here was conducted by the on-surface synthesis of aza-[5]triangulene, via a one-step annealing process on Au(111). Then scanning probe microscopy measurements were performed \cite{Lawrence_Lu_2023}. The dI/dV images presented in Fig \ref{fig3} (e) and (f) were obtained in the constant current mode with a CO tip.

In all three cases, we present the probability density of the lower lying state (denoted as $\Psi_{1\uparrow}$ in Fig. \ref{fig2}(c)) separately from the rest of the degenerate band (denoted as $\Psi_{2\uparrow}, \Psi_{3\uparrow}, \Psi_{4\uparrow}$ in Fig. \ref{fig2}(c)). One can note that all three methods yield similar results - the lower lying state is localized predominantly at the centers of the edges, with some non-zero contribution at the center of the quantum dot, while the other three states are localized entirely on the edges. 

We can now generalize our problem to an aza-[5]triangulene with an impurity localized on an arbitrary atom. Fig. \ref{fig4}(a) shows an analysis of the energy  $\Delta$ of the split-off level for all possible positions of the nitrogen atom on the majority sublattice. The red data points refer to the impurity localized on the edge of the quantum dot (red atoms in Fig. \ref{fig4}(c)), while the orange data points correspond to the bulk atoms, depicted also in orange in Fig. \ref{fig4} (c). Panel (b) shows the magnitude of the wave function coefficient $|b_n|^2$ corresponding to the site with impurity. The colors of the data points have been assigned analogically as in (a). One can notice that there is a clear correlation between the magnitude of the energy split and the wave function coefficient on a given atom. This leads us to the conclusion that positioning the impurity at a given site, and measuring the split between the first state and the rest of the degenerate shell, allows us to effectively probe the wave function localized on that site. One can notice, that both the split and the coefficient $b_n$ are larger for the atoms corresponding to the edges of the sample (red data points). This observation is in agreement with the wave function densities presented in Fig. \ref{fig3}, where we saw that most of the wave function density is indeed localized at the edges of the quantum dot.

We note, that moving the nitrogen impurity over the minority sublattice doesn't affect the structure of the degenerate shell, since the B-atom coefficients have no contribution to the zero-energy wave functions.


\section{Conclusions}
In this work, we showed how to probe the wave functions of a degenerate shell in a triangular graphene quantum dot with broken sublattice symmetry, using a localized impurity. We demonstrated its applicability on the example of aza-triangulenes. The analytical solution for the degenerate states of an all-carbon triangulene, solutions of the singular eigenvalue problem, was used as a basis for a triangulene containing a nitrogen impurity. The structure of the zero energy band in the presence of an impurity was predicted in terms of a one level peeling off. We showed that the impurity allows the probing of the wave functions of a degenerate shell on a carbon site where the impurity is located. Finally, we confirmed our predictions by the successful comparison with the tight-binding and \textit{ab-initio} calculation as well as with the experiment.

\section*{Acknowledgments}

A.W.R. thanks P. Potasz for helpful discussions. A.W.R., D.M. and P.H. acknowledge support by NSERC Discovery Grant No. RGPIN 2019-05714, the QSP-078 project of the Quantum Sensors Program at the National Research Council of Canada, the HTSN-341 project of the High-Throughput \& Secure Networks Challenge Program at the National Research Council of Canada, University of Ottawa Research Chair in Quantum Theory of Materials, Nanostructures, and Devices, and computing resources at the Digital Research Alliance Canada \cite{alliance_canada}.
J.Lawrence acknowledges the support from the Agency for Science, Technology and Research (A*STAR) Advanced Manufacturing \& Engineering (AME) Young Individual Research Grant (YIRG) M23M7c0121.
J.Lu acknowledges the support from Ministry of Education, Singapore Tier 2 grant ( MOE-T2EP10123-0004).


%
%
\bibliography{bibliography}
\end{document}